# UNRAVELING THE NUANCES OF AI ACCOUNTABILITY: A SYNTHESIS OF DIMENSIONS ACROSS DISCIPLINES

*Completed Research Paper*

Long Hoang Nguyen, Karlsruhe Institute of Technology, Karlsruhe, Germany, long.nguyen@kit.edu

Sebastian Lins, Karlsruhe Institute of Technology, Karlsruhe, Germany, sebastian.lins@kit.edu

Maximilian Renner, Karlsruhe Institute of Technology, Karlsruhe, Germany, maximilian.renner@kit.edu

Ali Sunyaev, Karlsruhe Institute of Technology, Karlsruhe, Germany, sunyaev@kit.edu

## Abstract

*The widespread diffusion of Artificial Intelligence (AI)-based systems offers many opportunities to contribute to the well-being of individuals and the advancement of economies and societies. This diffusion is, however, closely accompanied by public scandals causing harm to individuals, markets, or society, and leading to the increasing importance of accountability. AI accountability itself faces conceptual ambiguity, with research scattered across multiple disciplines. To address these issues, we review current research across multiple disciplines and identify key dimensions of accountability in the context of AI. We reveal six themes with 13 corresponding dimensions and additional accountability facilitators that future research can utilize to specify accountability scenarios in the context of AI-based systems.*

## 1 Introduction

AI has experienced a surge in popularity. For organizations and society, the utilization can be beneficial in several ways (Thiebes et al., 2021), such as significantly increasing organizational performance (e.g., automating repetitive tasks; Wamba-Taguimdje et al., 2020) or offering innovative AI-based systems (e.g., autonomous vehicles; Hengstler et al., 2016). Large technology organizations are participating in an AI race to establish themselves as the dominant force in offering AI technologies and having a large market share (Brecker et al., 2023). Along with the advancing AI adoption across industries, scandals are rising and reveal various issues in contemporary AI-based systems. For example, the risk of discrimination by AI-based systems such as chatbots or image labeling algorithms has caused public attention to potential issues associated with pervasive AI use (Schmidt et al., 2020). Those scandals show that AI-based systems can inflict minor to serious or even lethal harms that are unintentional and sometimes even intentional (e.g., hidden information, intentions, and behavior; Wieringa, 2020). Indeed, users of AI-based systems face manifold challenges today, including ethical, privacy, and cybersecurity risks (Du and Xie, 2021). To address these challenges, it becomes increasingly important to know who is accountable for harmful intentional or unintentional consequences that result from AI usage. In other words, there is a need to establish AI accountability to guarantee that actors justify their actions and respond to interrogations, and victims are compensated and perpetrators punished accordingly, among others (Bovens, 2007).

Nevertheless, achieving AI accountability is still challenging for legislative administrations, organizations, and individuals. Legislative administrations face a scarcity of legal and ethical frameworks guiding accountability processes. For example, ongoing debates elaborate on whether AI-based systems can be held accountable, even though they lack self-awareness (Deibel, 2021; Naja et al., 2022). Organizations face the opaque nature of AI-based systems (Habli et al., 2020), which results in the necessity of many stakeholders involved in the design, development, and operation of AI-based systems (e.g., developers, managers, companies), leading to a divergent understanding of who can be held accountable (i.e., problem of many hands; Webb et al., 2020). Individuals similarly face uncertainty about whether they can be held accountable using AI-based systems. Today, practice still faces ambiguity on how to handle and ensure AI accountability, while the number of AI-based systems is increasing every day.

Given these challenges for practice, an ever-increasing number of researchers from various disciplines examine, among others, how to foster accountability and predict its consequences on individual and organizational behavior. For example, computer scientists propose new technical frameworks that support AI accountability (e.g., Naja et al., 2022), whereas legal scholars often discuss the applicability of legal frameworks to enforce accountability and issue sanctions (e.g., Kaminski, 2018). Related information systems (IS) research has particularly analyzed how individual's perceptions of being accountable impact their intentions and behaviors (e.g., Vance et al., 2013, 2015).

However, these valuable research endeavors strengthen the problem that research on AI accountability is highly scattered across multiple disciplines, with each discipline having a different or nuanced view of accountability. Since different disciplines have divergent understandings of AI accountability, there is difficulty in approaching accountability properly in the AI context. This problem leads to conceptual ambiguity on AI accountability (Kempton et al., 2023), hampering the comparison of research findings. This is especially problematic for IS research, as IS research is interdisciplinary and multifaceted in general and influenced by different research streams (e.g., social sciences, computer sciences, and law; Sarker et al., 2019). Consequently, researchers have called for further research to create a more unanimous understanding of AI accountability (e.g., Memarian and Doleck, 2023). Identifying the key dimensions of AI accountability that synthesize and represent knowledge from multiple disciplines can serve as a foundation to increase our understanding. Accordingly, we aim to answer the following research question (RQ): *What are the key dimensions of AI accountability?*

We conducted a descriptive literature review (Paré et al., 2015) that synthesizes the scattered research on AI accountability across several disciplines. Our analyses revealed six key themes of AI accountability, namely trigger, entity, situation, forum, criteria, and sanctions. We further divided these themes into 13 dimensions emphasizing the prevalent nuances around AI accountability. In addition, we uncover important accountability facilitators (e.g., system-related characteristics or features that are known to benefit accountability). Our study addresses calls for interdisciplinary syntheses by considering studies from multiple disciplines, including computer science, law and policy, and IS. We contribute to research by providing a set of combinable dimensions to capture accountability scenarios comprehensively, a common ground for research to approach and examine AI accountability, and a recent view on AI accountability by considering emerging accountability topics like algorithmic actors.

## 2 Background

### 2.1 AI accountability

Within this study, we follow a broad definition of AI as "the ability of a machine to perform cognitive functions that we associate with human minds, such as perceiving, reasoning, learning, interacting with the environment, problem-solving, decision-making, and even demonstrating creativity" (Rai et al., 2019, p. iii). Building on this general definition, we do not differentiate between different types of AI systems while we acknowledge that AI system diversity and contextual conditions shape the accountability process. Ensuring AI accountability becomes an increasingly important topic to address the existing uncertainties and challenges faced by organizations and users when adopting and using AI-based systems (e.g., ethical issues such as discriminating AI predictions). Both academia and practice consider accountability as key when handling AI-based systems, for example, expressed by prominent frameworks like FATE (i.e., Fairness, Accountability, Transparency, Ethics; Memarian and Doleck, 2023).

In its essence, accountability is referred to as "a relationship between an actor and a forum, in which the actor has an obligation to explain and to justify his or her conduct, the forum can pose questions and pass judgment, and the actor may face consequences." (Bovens, 2007, p. 450). Though, this definition mainly takes on a legal perspective and could, therefore, not consider all aspects that emerge in the context of designing and developing AI-based systems. Consequently, Wieringa (2020) contextualized Bovens (2007) prominent definition of accountability. She summarizes AI accountability as creating an account for a socio-technical AI system that involves multiple actors (e.g., decision-makers, developers, users) who have an obligation to explain and justify their use, design, or decisions concerning the AI system and subsequent consequences. These actors may be held accountable by various types of fora (e.g., external to the organization) for particular aspects of the system (e.g., the software code) or the entirety of the system. This study builds on this contextualized definition by Wieringa (2020), while considering its roots in the valuable work of Bovens (2007) and complementary accountability research (e.g., Day and Klein (1987)). Reflecting these definitions, we argue that an accountability process covers three important phases: (1) an information phase, in which an actor gives information on his/her actions related to the AI system to the forum; (2) the deliberation and discussion of the forum; and (3) imposing consequences on the actor by the forum in case the actor is held accountable (Brandsma and Schillemans, 2012). An accountability process is embedded in a specific context and its conditions (e.g., a specific type of AI system). The context then shapes the manifestation of the accountability process, for example, requiring different fora.

Notably, accountability is related but different to prominent AI concepts (i.e., transparency, explainability, auditability). For example, technical transparency of the (inner) workings of the AI system is helpful but does not elaborate on the responsible person and the decision-making process of certain actors (Wieringa, 2020). While transparency is passive (i.e., 'see for yourself how it works') and can be helpful during the accountability process, holding an actor accountable requires a more active and involved stance (i.e., 'let me tell you how it works, and why'; Wieringa, 2020). Despite its interrelatedness, achieving accountability in the context of AI is challenging. For instance, there are often multiple actors involved (e.g., developers, users, companies). This creates ambiguity when trying to identify the main accountable actor, leading to the problem of many hands (Webb et al., 2020). An ever-increasing number of research has, therefore, started to examine and resolve prevalent accountability gaps.

### 2.2 Related research

There is a large body of research on AI accountability scattered across diverse disciplines, predominantly in computer science, law and policy, as well as IS. Each discipline takes a different perspective on accountability, leading to diverging research endeavors (Table 1).

| Disciplines | Exemplary related research endeavors | Exemplary studies |
| --- | --- | --- |
| Computer Science | Introduction of technical frameworks fostering the development and governance of accountable AI-based systems. | Naja et al., 2022; Sokol et al., 2022 |
| Law and Policy | Analyses and comparison of existing legal frameworks, or introduction of new policing approaches to achieve accountability. | Kaminski, 2018; Mökander et al., 2022; Oswald, 2022 |
| Information Systems | Observations of the impact that being accountable has on individuals' perceptions and behaviors. | León et al., 2021; Shklovski and Némethy, 2022; Vance et al., 2013, 2015 |

*Table 1. Exemplary research endeavors of related disciplines.*

Studies rooted in computer science primarily deal with introducing technical frameworks and means fostering the development and governance of accountable AI-based systems. For instance, technical frameworks such as knowledge graphs propose an aid to clarify relevant accountability information for all stakeholders (Naja et al., 2022). Knowledge graphs can be beneficial, as they provide key information in a structured format understandable by both humans and machines (e.g., on the creators of an AI-based system and its intended use cases). Similarly, there are proposals for comprehensive toolboxes that can automate large parts of the AI auditing process. These toolboxes, for example, apply metrics, such as data density, to judge the accountability of AI-based systems (Sokol et al., 2022). Data density can be treated as a proxy for prediction confidence and thus be an indicator of its robustness, which in turn contributes positively to accountability (Sokol et al., 2022).

Unlike in computer science, research in the fields of law and policy addresses the issue of AI accountability from a legal perspective. Manifold studies within this domain focus on analyzing and comparing existing legal frameworks. To cite an instance, the European General Data Protection Regulation (GDPR) and its relevancy for AI accountability is a common subject of research discussion (Kaminski, 2018). Comparably, the recently emerging European AI Act and the United States Algorithmic Accountability Act (US AAA) are frequently being contrasted as part of legal analyses (Mökander et al., 2022). Nonetheless, studies from the field of law and policy do not only reflect upon existing legal approaches but also include novel policing approaches as a research endeavor. For instance, a three-pillar approach is suggested to guide future policing (Oswald, 2022). According to this approach, achieving accountability from a legal perspective entails (1) the application of relevant law while (2) considering current ethical and scientific standards for (3) all people across societal layers.

Finally, there are research streams from the IS field that put less emphasis on surrounding technical and legal frameworks but rather on individuals facing accountability. One of the many key research interests of IS research is the impact of accountability on the perception and behavior of individuals. Accordingly, there are, for example, first investigations on how AI developers feel when being confronted with ethical concerns related to the applications they develop (Shklovski and Némethy, 2022). There are also investigations on the effect of accountability pressure on individuals' behavior. For instance, while facing accountability pressure, individuals are more likely to share resources with AI partners, take more time to make decisions, and perform worse in tasks with better-performing AI partners (León et al., 2021).

Summarizing all these research endeavors, they provide a valuable and solid foundation for examining AI accountability within each respective field. However, conceptual ambiguity emerges because insights are scattered across multiple disciplines (Kempton et al., 2023), with each discipline having a different perspective on accountability. We, therefore, lack a common understanding of the key dimensions of AI accountability. This, in turn, is problematic because research findings may not be compared and integrated across disciplines. Researchers and practitioners may oversee key facets and potential pitfalls when dealing with AI accountability, even if they have already been revealed in related research. Wieringa (2020) similarly notes the risk of "miscommunication between disciplines" (p.10). Consequently, there are recent calls for further research to work towards a more unanimous understanding across disciplines (e.g., Memarian and Doleck, 2023; Wieringa, 2020). Synthesizing key dimensions of AI accountability across disciplines could serve as a foundation to achieve that goal and provide a common set of dimensions that researchers should consider when examining AI accountability.

There are already a few recent studies conducting literature reviews to mitigate the prevalent conceptual ambiguity. Kempton et al. (2023), for instance, review multiple definitions of accountability and discuss how extant literature informs the management of AI. They revealed that "almost one out of two […] reviewed papers use the term accountability without defining what the term means" (p. 5) and recommend that IS researchers should take a more holistic view on AI accountability, considering its diverse dimensions. To approach this conceptual ambiguity on a deeper level, Wieringa (2020) elaborated on the widely accepted definition of accountability by Bovens (2007) as a foundation for a contextualization considering AI specifics. Her study offers a thorough elaboration on AI accountability based on Bovens' accountability theory but is limited to literature until 2018, and hence, misses recent developments in the field of AI. Moreover, there are further theoretical frameworks on accountability, which can be utilized complementary to Bovens' definition (e.g., Day and Klein, 1987). To answer recent calls for an interdisciplinary conceptualization, we set out to synthesize research on accountability from multiple disciplines.

## 3 Method

We conducted a descriptive literature review (Paré et al., 2015) on AI accountability. Our main goal was to identify dimensions of accountability in the context of AI while applying established guidelines for literature reviews (Brocke et al., 2009; Webster and Watson, 2002).

### 3.1 Literature search

We determined the search string to reveal articles that deal with AI and accountability. We decided on a more general search string, as adding AI-related keywords like deep learning and accountability-related keywords like ethics did not yield more relevant results. We applied the search string *(Artificial Intelligence OR AI) AND (Accountab\*)* in five scientific databases, selected for their access to high-quality, peer-reviewed articles in various disciplines: EBSCOhost, ProQuest, IEEE Xplore, AIS Electronic Library, and ScienceDirect. We limited our search to title, abstract, and author keywords, yielding a total of 880 potentially relevant articles as of July 3, 2023.

We conducted a relevancy check in two stages. First, all 880 articles were assessed based on their title and abstract. We applied inclusion (e.g., discussing dimensions of accountability) and exclusion criteria, leading to the exclusion of 775 articles marked as duplicates (149), not in English (22), grey literature (30), off-topic (146), and AI studies not including accountability as a key aspect (e.g., only stating that ensuring AI accountability is important; 428). Second, the remaining 105 potentially relevant articles were analyzed in their entirety. 67 articles remained for analysis after excluding 38 articles dealing with other aspects unrelated to AI accountability dimensions, such as suggested technical implementations of an exemplary accountable AI-based system (e.g., Wahde and Virgolin, 2023).

### 3.2 Literature analysis

We applied thematic analysis to our final set of 67 articles as a structured approach to identify dimensions of AI accountability. The thematic analysis includes six steps: familiarizing yourself with the data, generating initial codes, searching for themes, reviewing themes, naming and defining themes, and lastly producing the report (Braun and Clarke, 2006).

During data familiarization, we took notes of each article's discipline, research focus, and article type, among others, to get an initial understanding. Through this step, we noticed that the 67 articles were spread across the disciplines of IS, computer science, law and policy, ethics, and social science. Most of the articles focus on AI ethics and governance in general. There were, however, also articles touching on, for example, specific approaches for accountable AI-based systems in healthcare and education.

For generating initial codes, we started by reading the full text of the articles and assigned initial codes to relevant text passages providing information on AI accountability dimensions. For instance, the text segment "The updated Algorithmic Accountability Act of 2022 would require impact assessments when companies are using automated systems to make critical decisions [...]" (Oduro et al., 2022, p. 3) was coded as 'Accountability Laws'. This coding process resulted in 109 codes assigned to 584 text segments. We iteratively refined the initial codes by merging too narrow and splitting too broad codes. For example, the codes 'AI Programmers' and 'Software Engineers' were aggregated to 'AI Developers'.

During the searching for themes step, we analyzed our initial codes and corresponding text segments to identify emerging themes. After the first round of analysis, we managed to identify 20 initial theme candidates. Those theme candidates included, among others, 'Individuals' (e.g., AI developers and AI users) and 'Organizations' (e.g., companies and regulatory bodies) that can be held accountable. After identifying an initial set of themes, we applied Patton's (2014) criteria of internal homogeneity and external heterogeneity. This guaranteed that the data within themes fits together meaningfully and that themes are distinct from each other.

In reviewing themes, we utilized the classification of general elements within accountability processes by Day and Klein (1987) to guide the entire consolidation process. We particularly chose this classification since it is established across the accountability literature, used in many disciplines, captures the key aspects of the widely applied definition of accountability provided by Bovens (2007), and supports us in aggregating our results and achieving higher levels of abstraction. The classification by Day and Klein (1987) comprises six interrelated elements of accountability that we used to categorize our final themes: trigger, entity, situation, forum, criteria, and sanctions. Comparing our themes to these elements helped us categorize 13 themes. For example, we assigned our themes 'Individuals', 'Organizations', and 'Algorithmic Actors' to the element 'Entity', summarizing entities that can be held accountable. Ultimately, our analyses yielded six higher-level themes comprising 13 distinct accountability dimensions (Table 2). These themes are embedded in the accountability process and are therefore interrelated. For example, a forum may impose different sanctions depending on specific triggers and the situation.

While applying thematic analysis and comparing our findings, we also identified three themes of 'Accountability Facilitators' that were frequently discussed in the literature. These facilitators include various factors, practices, and conditions that contribute positively towards achieving accountable AI-based systems and support the accountability process. The three underlying themes of facilitators are namely: (1) governance mechanisms (i.e., mechanisms for internal and external oversight of organizations like reports and audits), (2) system properties (i.e., system-related characteristics or features that are known to benefit accountability), and (3) social features (i.e., people-related characteristics or features that are mostly non-analytical). We present these facilitators in Section 5.1 in more detail as they were predominantly discussed and emphasized across disciplines (i.e., 159 codes across 52 articles).

In defining and naming themes, we based our definitions on the classification provided by Day and Klein (1987). All themes are reported in Section 4, which also addresses the final step in the process: producing the report.

# 4    Key Dimensions of AI Accountability

| Theme | # Coded / # Articles | Dimension | Description | Example Sources |
|---|---|---|---|---|
| Trigger | 52 / 36 | Errors and biases | AI-based systems performing improperly or not as expected, thus not meeting standards. | Bagave et al., 2022; Omeiza et al., 2022 |
| | | AI consequences | Consequences of using AI-based systems, in this case, mainly negative (i.e., harms). | Fletcher and Le, 2021; Ozanne et al., 2022 |
| | | Violations | Contract and law violations, often linked to unethical decisions and tortious acts. | Hammond, 2014; McGregor et al., 2019 |
| Entity | 182 / 51 | Individuals | Natural and legal persons, including the creators and users of AI-based systems. | Johnson, 2022; Kiseleva, 2020 |
| | | Organizations | Collective natural and legal persons, including private and public sector bodies. | Banteka, 2020; Singh et al., 2019 |
| | | Algorithmic actors | Direct source of the harm, currently not being recognized as a natural or legal person. | Khan and Vice, 2022; Webb et al., 2020 |
| Situation | 29 / 19 | AI lifecycle | Entire AI lifecycle, including conception, design, development, deployment, and use. | Naja et al., 2022; Raja and Zhou, 2023 |
| Forum | 55 / 27 | Individuals | Users and persons negatively impacted by the usage of AI-based systems (i.e., victims). | Oduro et al., 2022; Sanderson et al., 2023 |
| | | Organizations | Operators of AI-based systems and traditional vs. purpose-built regulatory bodies. | Busuioc, 2021; Percy et al., 2021 |
| Criteria | 69 / 41 | Laws | Enforceable requirements aiming to hold entities accountable without restricting use. | Kaminski, 2018; Mökander et al., 2022 |
| | | Standards | Primarily ethical recommendations that lack legal power to ensure enforceability. | Katyal, 2019; Smith, 2020 |
| Sanctions | 38 / 24 | Punitive measures | Imposed when retrospective explanations or justifications for misconduct are inadequate. | Bickley and Torgler, 2022; Busuioc et al., 2022; |
| | | Redress | Reimbursement of affected victims instead of penalizing entities to prevent future harm. | Fukuda-Parr and Gibbons, 2021 |

*Table 2.    Summary of AI accountability dimensions.*

## 4.1 Trigger

An accountability process is typically evoked by a trigger that is an event (Day and Klein, 1987). In the context of AI, events include disclosing errors and biases (e.g., in training data), (negative) consequences of using AI-based systems, or contract and law violations.

Generally, when errors or biases occur, the AI-based system is not performing properly and according to expectations (Bagave et al., 2022), thus failing to meet standards (Omeiza et al., 2022). Errors and biases can lead to issues like discrimination against individuals or groups in society, which then triggers an AI accountability process (Bannister et al., 2020; Katyal, 2019). Prominent examples are false diagnostic treatment decisions in healthcare (Ahmad et al., 2020; Murphy et al., 2021) and racial bias during college admissions (Percy et al., 2021). Potential origins for this adverse behavior are related, for example, to the datasets used for model training being corrupted (Santoni de Sio and Mecacci, 2021) or suffering from sample-size disparity, which leads to a lack of representativeness (Sokol et al., 2022; Webb et al., 2020). Another cause for errors and biases that is unrelated to data could be the improper use of technology due to lack of training (Ahmad et al., 2020).

Further accountability triggers relate to the consequences of using AI-based systems. While there can be positive consequences of AI use, prior literature mainly examines negative events triggering accountability processes, namely the harm caused to individuals, markets, or society (Fletcher and Le, 2021). These harms can be unintended (Ozanne et al., 2022), tend to have a grievous nature (Tóth et al., 2022), and directly affect the human life and livelihoods (Bickley and Torgler, 2022). In the case of healthcare, false predictions during blood tests can lead to wrong treatments, which lead to, for instance, overdoses (Habli et al., 2020). In other cases, they can also violate citizens' rights (Donahoe and Metzger, 2019).

The third stream of triggers is related to contract and law violations. Generally, the accountability process can be triggered when a violation of the legislation or law occurs. These violations are often linked to unethical decisions and tortious acts (Deibel, 2021; Shklovski and Némethy, 2022) and can range from human rights violations (McGregor et al., 2019) to extreme cases of war crimes committed with AI-based systems (Hammond, 2014). Similarly, accountability processes can be triggered due to contract violations. Accountability in contracts relates to identifying contract breaches and the responsible party that owes remedy to the affected party (Singh et al., 2019).

## 4.2 Entity

Accountability processes include one or multiple entities that are accountable or held accountable (Day and Klein, 1987). Entities refer to individuals (Maas, 2022; Memarian and Doleck, 2023) and organizations (Johnson, 2022) but can also refer to algorithmic actors in the AI context (Ozanne et al., 2022).

Entities on an individual level are natural and legal persons (Singh et al., 2019). These persons include creators of AI-based systems (e.g., developers, designers; Bagave et al., 2022; Johnson, 2022) and users that rely on decisions made by AI-based systems (e.g., doctors, nurses; Choudhury and Urena, 2022; Kiseleva, 2020). Generally, any person can be subject to accountability, regardless of their status within organizations (Oswald, 2022). However, the extent of accountability might differ depending on factors like the individual ability to evaluate recommendations from AI-based systems (Kiseleva, 2020).

Like individual entities, organizations can be considered both natural and legal persons since persons resemble artificial entities under federal law (Banteka, 2020; Singh et al., 2019). The main difference to individuals is that organizations are collective entities (Johnson, 2022). Organizations incorporate private sector companies (e.g., corporations that develop AI-based systems or supply data; Memarian and Doleck, 2023; Sanderson et al., 2023) and public sector bodies (e.g., local and federal governments; Busuioc, 2021). Public sector bodies often outsource the development of AI-based systems to private sector companies, which lack political accountability (Santoni de Sio and Mecacci, 2021). Regardless, accountability must be considered since companies can still be accountable when acting under the government (Crawford and Schultz, 2019). The more inscrutable an algorithm is designed, the more accountability falls onto the private sector company developing the algorithm (Martin, 2018).

The last emerging entity tries to attach accountability to the direct source of the harm, the algorithmic actor itself. As AI-based systems increasingly perform highly critical tasks, they might also be considered accountable (Khan and Vice, 2022; Webb et al., 2020). However, since algorithms cannot be considered natural and legal persons, accountability can only be ensured by granting legal personhoods (Banteka, 2020). As algorithms currently do not have a legal personality, they cannot be considered accountable under existing technological and legal standards (Deibel, 2021; Fletcher and Le, 2021). Therefore, accountability generally falls back onto the individuals and organizations that created or influenced the behaviors of AI-based systems (Naja et al., 2022). In some cases, it is even suggested that both humans and algorithms share accountability when issues occur (Memarian and Doleck, 2023).

### 4.3 Situation

Events as triggers of accountability processes are tied to specific situations in which entities perform certain actions (Day and Klein, 1987). In the context of AI-based systems, these situations primarily revolve around the entire AI lifecycle, which is closely linked to the development process.

The development process itself only represents one fraction of the entire AI lifecycle, which also entails the phases before and after developing the AI-based system (Bagave et al., 2022). However, considering the entire AI lifecycle is crucial to achieving accountable AI-based systems (Maas, 2022; Naja et al., 2022). To do so, Raja and Zhou (2023), for instance, proposed a model capturing the entire AI lifecycle beyond the development phase. According to their model, the lifecycle starts at a conception phase, where the initial understanding of a business problem is created. This is followed by a design phase that aims to make the initial understanding unanimous for a clear definition of deliverables. The final understanding should lay out every underlying assumption to set forth potential ethical concerns and biases, which could be relevant for accountability. Thereafter, the phase of data acquisition and preparation starts, which amounts to the development of an AI model that is trained and evaluated. Factors that directly affect accountability here, for instance, include the way training data is handled. Sensitive data, in particular, requires special attention to avoid later accountability concerns. After the evaluation and, thus, the development is finished, the AI model gets deployed. The deployment phase also ensures that all regulatory requirements are met to foster accountability. Finally, there is a phase in which the system is in operation and simultaneously monitored, for instance, through impact assessments. A continuous impact assessment is especially important in the context of AI-based systems, as these systems continue to learn with data provided by users, which leads to blurred accountability boundaries between the original developers and the users (Sanderson et al., 2023). The lifecycle of AI-based systems from Raja and Zhou (2023) closes with the monitoring phase and starts again, beginning at the conception phase.

### 4.4 Forum

Corresponding to entities in accountability processes, there is an accountability forum to whom entities are accountable (Day and Klein, 1987). In general, accountability fora are audiences with the authority to reward and sanction the entities (Tóth et al., 2022). Though, accountability fora can also be affected parties that believe an entity is obligated to account to them. Any individual or organization fulfilling these criteria can resemble an accountability forum (Johnson, 2022).

When it comes to individuals, accountability fora mainly revolve around the actual users of AI-based systems and the individuals who are negatively impacted by the usage of AI-based systems (i.e., victims; Sanderson et al., 2023). Users particularly include customers who privately or commercially utilize AI-based systems offered by private organizations like Google (Donahoe and Metzger, 2019). As customers of these organizations, these users have the right to request and receive information about algorithmic decision-making pertaining to themselves. Due to that right, they resemble an accountability forum (Oduro et al., 2022). Victims, on the other hand, include individuals who are directly affected by triggers (e.g., errors and biases) that result from the use of AI-based systems by other individuals or organizations (Santoni de Sio and Mecacci, 2021). In the case of healthcare, for example, victims would be the patients, dependent on AI-supported decisions (Bagave et al., 2022). Similarly, victims in the defense context suffered from war crimes committed with AI-based systems (Hammond, 2014).

Accountability fora can vary greatly depending on the context. Therefore, an entity might be accountable to individual victims in one context while it is accountable to organizations in another context. Organizational fora can include, among others, operators of AI-based systems and regulatory bodies (Singh et al., 2019). Regulatory bodies resemble both traditional fora, such as courts and parliamentary committees, as well as purpose-built fora, such as ethics, standardization, and audit bodies (Busuioc, 2021). Purpose-built fora are created to govern either specific parts or the entirety of an AI-based system (Van den Homberg et al., 2020) and can be appointed by an entity itself (e.g., developer review boards) (Percy et al., 2021). A prominent example of an organizational forum is the Association for Computing Machinery (ACM) with its statements on ethics for AI developers (Johnson, 2022).

### 4.5 Criteria

Within accountability processes, a forum determines criteria that are applied to judge different triggers accurately. These criteria are generally closely tied to the current legislation and the current ethical or political standards (Day and Klein, 1987).

There are various attempts from legislation to address the AI accountability issue, with the EU GDPR being one of the first. The GDPR was introduced in 2018 and addresses AI accountability by providing a 'right to explanation' if personal data is processed. This right allows individuals to obtain meaningful information about the logic in automated decision-making (Kaminski, 2018). Moreover, the GDPR envisions the establishment of ethical review boards, which can further enhance AI accountability (Kaminski, 2018; Singh et al., 2019). On top of the GDPR, the EU introduced a first draft of the EU AI Act later in 2021, which finally passed in March 2024. This act is grounded in human rights and aims to ensure that AI-based systems are not discriminating based on gender, race, or other demographic traits (Oduro et al., 2022). Corresponding to the EU AI Act, the US proposed an updated version of their AAA from 2019 in 2022 (Khan and Vice, 2022; Oduro et al., 2022). Both the EU AI Act and the US AAA seek to establish governance infrastructures to hold entities accountable while not prohibiting the use of AI-based systems (Mökander et al., 2022). In addition to these well-known frameworks, there were manifold legal proposals published by US states (e.g., the California Automated Decision Systems Accountability Act of 2021) (Oduro et al., 2022). Some industries, like healthcare, also mainly adhere to existing legal frameworks tailored to their field (e.g., the Health Insurance Portability and Accountability Act - HIPAA) to address AI accountability (Kiseleva, 2020; Omar, 2020).

Not all professions have already established regulatorily enforced requirements like healthcare does. AI developers, for instance, mostly follow recommended ethical standards (Smith, 2020). One of the most prominent ethical standards was introduced by the ACM in 2017. According to the ACM, AI developers should, among others, ensure that there is awareness of possible biases, that there are mechanisms for individuals to question and address adverse effects, and that someone is held accountable for their use of algorithms (Johnson, 2022; Katyal, 2019). Similarly, the Institute of Electrical and Electronics Engineers (IEEE) published a report on 'Ethically Aligned Design' earlier in 2016. This report stated, for example, that all systems should embed human norms and values (Katyal, 2019). However, as standards primarily resemble recommendations, there is no legal power in place that enforces rules (Smith, 2020). Thus, governments, companies, and academic institutions constantly dedicate efforts to find universally applicable and impactful standards (Percy et al., 2021). As part of these efforts, the European Commission proposed the Trustworthy AI Guidelines in 2019. These guidelines, among others, recommend audits across the development process, including deployment and use (Bagave et al., 2022; Maas, 2022).

### 4.6 Sanctions

Depending on the case, sanctions may be imposed on an entity by a forum (Day and Klein, 1987). These sanctions can be aimed at directly punishing entities as well as providing redress for those adversely affected, including compensation or reparation (Bickley and Torgler, 2022; Busuioc, 2021).

Punitive measures are generally imposed when retrospective explanations or justifications for misconduct are inadequate (Busuioc et al., 2022). These punitive measures vary greatly depending on the severity of the misconduct and can range from facing criticism by the public, demotion or firing from a workplace, a fine, and, in extreme cases, a jail sentence (Bannister et al., 2020). Taking non-compliance with the GDPR for example, organizations must pay up to 20 million Euros, or in the case of large organizations, up to 4% of their annual turnover as a fine. Furthermore, regulators may impose corrective orders and prohibitions on processing, heavily restricting an organization (Singh et al., 2019). As for individual professionals like clinicians, a breach of any code of conduct could lead to getting fired or a prohibition from practicing (Smith, 2020).

Apart from directly punishing entities, providing timely and effective redress is a key aspect related to accountability (Fukuda-Parr and Gibbons, 2021). As opposed to punitive measures, redress has a strong foundation in human rights and additionally considers reimbursing the affected victims instead of only penalizing entities. A common goal of punishments and redress is to ensure that harms are not repeated in the future (McGregor et al., 2019). According to the United Nations Framework on Business and Human Rights, redress may include apologies, restitution, rehabilitation, financial or non-financial compensation as well as binding guarantees of non-repetition (Donahoe and Metzger, 2019).

# 5 Discussion

We conducted a descriptive literature review to synthesize recent research on AI accountability across multiple disciplines. To address the conceptual ambiguity related to AI accountability, we propose six key themes of AI accountability, including trigger, entity, situation, forum, criteria, and sanctions. We dissected these themes into 13 interrelated dimensions to emphasize the existing nuances around AI accountability while keeping a higher level of abstraction. Besides the 13 key dimensions, we also uncovered accountability facilitators that we will discuss in the following. To conduct an often-demanded interdisciplinary synthesis, we considered studies from manifold disciplines, most prominently computer science, law and policy, as well as IS.

## 5.1 Accountability facilitators

Accountability facilitators comprise factors, practices, and conditions that contribute positively towards achieving accountable AI-based systems. Facilitating aspects are of socio-technical nature and include but are not limited to organizational governance mechanisms, system properties, and social features.

Achieving accountability through governance typically entails appointing independent internal compliance officers, producing internal reports, and involving independent external oversight (i.e., third-party auditing instances; Kaminski, 2018). The extent of applying these governance mechanisms depends on the risk associated with the reviewed AI-based system. A high-risk system would require more resources as well as increased levels of monitoring after the deployment to ensure accountability (Bazoukis et al., 2022; Percy et al., 2021). To save resources, there are recent suggestions for automated oversight programs that can perform black-box audits and thus replace human oversight (Bickley and Torgler, 2022; Murphy et al., 2021). Nonetheless, most guidelines still emphasize human beings and their key role in AI accountability governance (Fukuda-Parr and Gibbons, 2021). This is particularly important to, for instance, counteract the automation bias, which leads to excessive trust or reliance on automated systems while underutilizing personal judgment (Engstrom and Ho, 2020).

To support governance mechanisms in the first place, AI-based systems should be developed with attention to system properties known to benefit accountability, such as auditability. Being auditable means providing the ability to inspect, review, and interrogate (Naja et al., 2022). This can include measures of reproducibility like record keeping and traceable logging (Bagave et al., 2022; Sanderson et al., 2023). Another system property relevant to accountability is transparency, which relates to the disclosure of information regarding, for instance, how AI-based systems are developed (McGregor et al., 2019). Transparency is especially important because it simplifies the process of identifying errors and biases for domain experts (Murphy et al., 2021; Sanderson et al., 2023). Manifold measures enabling transparency are proposed in research and practice, such as use case diagrams (Takeda et al., 2019) and source code disclosure (Engstrom and Ho, 2020). Though, even a high degree of transparency through access to the entire source code may not yield accountability (Katyal, 2019). This is because a singular system property may be necessary but is rarely sufficient to achieve accountable AI-based systems (Busuioc, 2021; Martin, 2018). Properties closely related to transparency like explainability and interpretability face similar difficulties due to their intertwined nature. Without transparency, it would become challenging to achieve meaningful explanations and interpretations. Explanations by themselves cannot be sufficient unless they are clearly understood by an observer. Thus, interpretability intends to capture both explainability and transparency to make an AI model understandable at multiple levels (Bagave et al., 2022). This difficulty already led to the emergence of simplicity as another related system property with the aim to enable accountability by reducing system complexity (Omar, 2020).

Complementary to system properties, there are facilitators like responsibility that primarily consider social features to enhance accountability. Responsibility can be referred to as the willingness of an entity to act in a transparent, fair, and equitable manner (Bagave et al., 2022). Despite responsibility being a non-analytical factor, it can still be used by regulatory bodies to draw conclusions regarding accountability (Busuioc, 2021). For that reason, a lack of responsibility can result in a lack of accountability (Santoni de Sio and Mecacci, 2021). Clinicians, for instance, should communicate any uncertainty regarding decisions made by AI-based systems as part of their responsibility. This mere act of responsible communication already enhances accountability (Smith, 2020). Similarly, developers would have to justify whether decisions made by their system are fair as part of their responsibility (Oduro et al., 2022). For organizations on the other hand, responsibility entails making sure that the staff handling AI-based systems is trained appropriately (Bazoukis et al., 2022). Without sufficient knowledge and training, it would be impossible for any professional to develop an awareness of the issues they are working on (Choudhury and Urena, 2022; Raja and Zhou, 2023). This lack of awareness in turn leads to a lack of responsibility and thus accountability (Santoni de Sio and Mecacci, 2021).

## 5.2 Open research challenges

While exploring the dimensions of AI accountability, we noticed that some dimensions are discussed in a controversial manner or are still fairly underexplored, offering avenues for further research. First, recent research engages in inconsistent discussions on whether the AI algorithm can be an entity (e.g., Khan and Vice, 2022; Webb et al., 2020). Studies mostly conclude that there is no way to hold algorithms accountable under current regulations because they still lack legal personhood (Deibel, 2021). However, in the context of self-learning AI, the question arises whether a human being is always at fault regardless of the circumstances and AI system type. Even though a human being might not be entirely at fault for a trigger, they would still be held accountable at present. Looking at the recently emerging criteria (e.g., the EU AI Act), the main point of interest remains the human being that is supposedly behind a trigger, and algorithmic entities are not addressed. Recent studies became aware of this issue and suggested that there might be ways for humans and algorithms to share accountability (Memarian and Doleck, 2023). Though, this approach is still largely underexplored. For IS research, we want to stress these controversies around entities to raise researchers' awareness about potential implications for study designs: Researchers should examine accountability scenarios where individuals (e.g., users) or organizations (e.g., the AI system provider) can be held accountable but not the AI system per se.

Second, our review reveals that the accountability situation is another underexplored theme. Most examined studies refer to the AI lifecycle from conception to use as the key point of interest. Nonetheless, research remains abstract and neglects single points within the AI lifecycle and largely neglects the diversity of AI systems. For example, accountability related to the usage phase after the AI-based system is deployed has attracted less research. As these systems tend to continue learning with new user data, there can be ambiguous accountability boundaries between the original developers and the users supplying the system with data (Sanderson et al., 2023). It thus becomes unclear whose behavior led to accountability triggers afterward, resulting in an accountability gap between developers' control and algorithms' behavior (Mittelstadt et al., 2016). IS research has recently started to examine this gap's consequences, for example, how much accountability developers self-attribute to them and how much accountability they perceive others attribute to them, creating intrapersonal perceptual accountability (in)congruence with unknown consequences for their job satisfaction (Schmidt et al., 2023). Taking a closer look at the AI lifecycle phases and different types of AI systems seems promising.

Finally, our findings reveal that potential sanctions are generally vaguely defined in current and emerging criteria. One aspect substantiating this vagueness is that accountability literature not only considers sanctions negatively (i.e., punitive measures) but also as rewards fostering accountability (e.g., Tóth et al., 2022). Nonetheless, the overwhelming majority of the literature refers to sanctions negatively, which is why we decided to follow the same notion. In the case of the GDPR, the values of financial penalties (e.g., 20 million Euros) for misconduct would mainly apply to larger organizations. Accountability cases sanctioning individuals are not part of legal statements yet, apart from minor remarks on facing criticism or firing from a workplace (Bannister et al., 2020). How this criticism would look in reality is, however, not mentioned. Redress faces similar issues and has no specific mention of, among others, the extent of rehabilitation for affected victims (Donahoe and Metzger, 2019). Interdisciplinary research engagement may counteract this knowledge deficit. For instance, law research may propose different sanctions, whereas IS research may then examine the impact of sanctions on entities' perceptions and behavior.

## 5.3 Implications for research

Our study has three key contributions for research. First, existing literature still faces conceptual ambiguity around AI accountability (Kempton et al., 2023). We reviewed 67 research articles and applied the classification from Day and Klein (1987) to contextualize AI accountability. Our findings reveal six key themes and additional accountability facilitators. We describe key manifestations of each theme by highlighting 13 corresponding dimensions that can be used by research as a theoretical basis to systematically examine accountability processes in-depth. Whereas research articles frequently underline the importance of AI accountability, less research has delineated the accountability process and its manifestation when examining accountability cases. With our findings, we inform research by providing a set of dimensions that can be combined to describe in detail and delineate a (class of) accountability scenarios. This becomes particularly important when contemplating the varied expressions of accountability, as underscored by our themes and their corresponding dimensions. For example, IS researchers investigating the repercussions of AI accountability mechanisms on individuals' perceptions and behavior can leverage our accountability dimensions to delineate precise accountability scenarios. Researchers can then describe the contextual conditions shaping the accountability scenario, including its process, the specific AI system, and the actors involved. This can assist them in interpreting their findings, pinpointing potential boundary conditions, and engaging in discussions regarding the generalizability of their results. In addition, our discussion on accountability facilitators can guide future research to select and examine possible antecedents of accountability (e.g., transparency, auditability) and to better elaborate on the differences and relationships between accountability and related constructs.

Second, existing literature regarding AI accountability is scattered across disciplines, with each discipline having an isolated point of view on accountability as a concept, risking a "miscommunication between disciplines" (Wieringa, 2020, p. 10). For example, computer science may not account for legal aspects of AI accountability, whereas law may lack depth on technical concepts. In this study, we aggregated literature across disciplines related to and including IS, most prominently computer science, law and policy. Our six themes and corresponding accountability dimensions provide a common ground to approach and examine AI accountability. We thereby also answer prior research calls for an interdisciplinary synthesis (e.g., Memarian and Doleck, 2023; Wieringa, 2020).

Finally, we acknowledge that recent studies (e.g., Wieringa, 2020) already started attempts to achieve contextualization (i.e., by also relying on Bovens (2007) accountability definition). Nevertheless, we believe that accountability literature offers a rich theoretical repertoire to consider. Thus, we openly extracted and synthesized various dimensions, then organized them systematically. Furthermore, existing reviews on AI accountability have overlooked recently emerging accountability topics, such as controversies surrounding algorithmic actors. Given the significant surge in the popularity and advancements of AI in recent years, our review offers a more up-to-date perspective on AI accountability and its pivotal dimensions, spanning diverse disciplines.

## 5.4 Implications for practice

Our findings can be utilized by practice to improve the understanding of what AI accountability processes entail. For individuals, we raise their awareness that they can be held accountable regardless of their individual status. Users, in particular, should not rely on the assumption that creators of AI-based systems are the main culprits and are always held accountable. Though, the level of expertise regarding these systems might impact the extent of accountability (Kiseleva, 2020). Besides, other revealed aspects like criteria and sanctions would especially be critical to consider for organizations of any size. Without considering current regulations and penalties for misconduct, the drawbacks of integrating a new AI-based system could quickly outweigh its benefits. On top of criteria and sanctions, our discussion on accountability facilitators helps consider not only the development of the system itself and technical means but also surrounding factors like governance and social features that foster accountability.

Like entities that can become aware of their accountability responsibilities with our results, victims and users, in general, obtain awareness about resembling an accountability forum (Sanderson et al., 2023). We want to emphasize that AI accountability enables them to hold any entity accountable for a perceived rights violation. Specifically, users may request information about the algorithmic decision-making process that is directly affecting them (Oduro et al., 2022). Apart from this distinct forum of individuals affected by AI-based systems, traditional organizational fora can also benefit from our findings. Among others, regulatory bodies and policymakers working on regulating AI-based systems can use our AI accountability dimensions to mitigate ambiguities regarding accountability processes.

## 5.5 Limitations & future work

Our study is subject to limitations that simultaneously pave the way for future studies to address. First, there is a set of limitations regarding literature reviews as the chosen methodology. We captured a broad set of 67 papers, which can, regardless, lack saturation. This is particularly apparent in themes that require further research as discussed in Section 5.2. Future studies may engage in detailed discussions with practitioners and researchers (e.g., conducting interviews), helping to gain rich descriptions and insights into each accountability dimension. Similarly, our findings emerged from the literature and still lack empirical validation. Future studies may use qualitative or quantitative research methods to validate and extend our findings. We also structured our themes based on a categorization by Day and Klein (1987), while other studies (e.g., Wieringa, 2020) focus on the complementary definition by Bovens (2007). We see avenues for future research to compare and integrate different established frameworks from accountability theory to offer a more holistic perspective on AI accountability. We also see merits in examining the relationships between different AI accountability dimensions. For example, a certain situation may require a different forum that relies on specific criteria.

Apart from methodology-related limitations, our study may be subject to content-related limitations. For example, when looking at entities, the literature mostly discussed that organizations are responsible for developing AI-based systems themselves. However, today's AI development is far more complex, involving cloud-based AI platforms that automatically generate models from data (Lins et al., 2021) and manifold open-source communities that might also be held accountable. We also acknowledge that different types of AI systems and contextual conditions shape the accountability process and the manifestations of our dimensions. We encourage researchers to engage with accountability and reflect on how AI's specifics hamper the accountability process.

## 6 Conclusion

Fostering accountability when developing and using AI-based systems gains increasing importance due to the widespread diffusion of AI-based systems and public scandals. However, we still face conceptual ambiguity, particularly since research is scattered across different disciplines having each their unique perspective on accountability. We reviewed extant research to uncover key dimensions of accountability. We reveal six themes and 13 corresponding dimensions that can be used by researchers to better delineate accountability scenarios in the context of AI-based systems.